\documentclass[aps,pre,twocolumn,superscriptaddress,showkeys,longbibliography]{revtex4-2}%
\usepackage{amsfonts}
\usepackage{amsmath}
\usepackage{amssymb}
\usepackage{graphicx}
\usepackage{color}
\usepackage{physics}
\usepackage{float}
\usepackage{mathtools}
\usepackage{algorithm}
\usepackage{algpseudocode}
\usepackage{wrapfig}
\usepackage{mathbbol}
\usepackage{upgreek}
\usepackage[normalem]{ulem}

\usepackage[colorinlistoftodos]{todonotes}
\usepackage[colorlinks=true, allcolors=blue]{hyperref}
\usepackage{mwe}
\usepackage{epstopdf}
\usepackage[doipre={DOI:~}]{uri}

\begin{document}
\title{Solid-State Dewetting of Polycrystalline Thin Films: a Phase Field Approach}

\author{Paul Hoffrogge}
\thanks{These authors contributed equally to this work.}
\affiliation{Institute of Nanotechnology—Microstructure Simulations (INT-MS), Karlsruhe Institute of Technology, Hermann-von-Helmholtz-Platz 1, 76344}
\affiliation{Institute for Digital Materials Science (IDM), Karlsruhe University of Applied Sciences, Moltkestrasse 30, 76133 Karlsruhe, Germany}
\author{Nils Becker}
\thanks{These authors contributed equally to this work.}
\affiliation{Institute for Applied Materials (IAM-MMS), Karlsruhe Institute of Technology, Strasse am Forum 7, 76131 Karlsruhe, Germany}
\author{Daniel Schneider}
\affiliation{Institute of Nanotechnology—Microstructure Simulations (INT-MS), Karlsruhe Institute of Technology, Hermann-von-Helmholtz-Platz 1, 76344}
\affiliation{Institute for Digital Materials Science (IDM), Karlsruhe University of Applied Sciences, Moltkestrasse 30, 76133 Karlsruhe, Germany}
\affiliation{Institute for Applied Materials (IAM-MMS), Karlsruhe Institute of Technology, Strasse am Forum 7, 76131 Karlsruhe, Germany}
\author{Britta Nestler}
\affiliation{Institute of Nanotechnology—Microstructure Simulations (INT-MS), Karlsruhe Institute of Technology, Hermann-von-Helmholtz-Platz 1, 76344}
\affiliation{Institute for Digital Materials Science (IDM), Karlsruhe University of Applied Sciences, Moltkestrasse 30, 76133 Karlsruhe, Germany}
\affiliation{Institute for Applied Materials (IAM-MMS), Karlsruhe Institute of Technology, Strasse am Forum 7, 76131 Karlsruhe, Germany}
\author{Axel Voigt}
\affiliation{Institute  of Scientific Computing,  TU  Dresden,  01062  Dresden,  Germany}
\affiliation{Dresden Center for Computational Materials Science (DCMS), TU Dresden, 01062 Dresden, Germany}
\author{Marco Salvalaglio}
\email{marco.salvalaglio@tu-dresden.de}
\affiliation{Institute  of Scientific Computing,  TU  Dresden,  01062  Dresden,  Germany}
\affiliation{Dresden Center for Computational Materials Science (DCMS), TU Dresden, 01062 Dresden, Germany}

\begin{abstract}
Solid-state dewetting is the process by which thin solid films break up and retract on a substrate, forming nanostructures. While dewetting of single-crystalline films is understood as a surface-energy-driven process mediated by surface diffusion, polycrystalline films exhibit additional complexity due to the presence of grain boundaries. Most theoretical and computational studies have focused on single-crystalline dewetting. Here, we present the application of the grandpotential multi-phase-field model to the dewetting of thin polycrystalline films in three dimensions, reproducing the key phenomenology of this process. By considering isotropic interface/surface energy, we illustrate its consistency with predictions based on energetic arguments and the morphological evolution towards equilibrium. We also provide novel analytical criteria for the onset of three-dimensional dewetting, serving as fundamental theoretical benchmarks, and highlight the critical role of triple junctions. Moreover, we unveil the dewetting behavior of polycrystalline patches, extending the scenarios of their single-crystalline counterparts.
\end{abstract}


\maketitle

Thin crystalline films are metastable due to their high surface-to-volume ratio. At high temperatures below melting, they typically break up and form three-dimensional agglomerates in a process known as solid-state dewetting (SSD) \cite{ThompsonARMR2012}. While detrimental to planar architectures, this phenomenon offers a unique kinetic pathway for the self-assembly of complex nanostructures, as clearly demonstrated for single-crystalline thin films. \cite{YeADVMAT2011,LeroySSR2016,NaffoutiSCIADV2017,BollaniNC2019}.
In these systems, the dynamics are predominantly governed by surface diffusion 
\cite{ThompsonARMR2012}. 

Several technologically relevant thin films are polycrystalline \cite{birkmire1997polycrystalline,kazmerski2012polycrystalline}: they consist of crystalline domains (grains) with varying crystallographic orientations, separated by interfaces (grain boundaries, GBs). Their morphological evolution is influenced not only by surface diffusion, as in single crystals, but also by additional effects associated with GBs \cite{MullinsJAP1957,Mullins1958,Srolovitz1986,Kaplan2013}. Arguably, the most relevant one is that surface grooves develop at GBs. Moreover, GBs migrate through a mechanism resembling attachment-detachment kinetics, with a velocity governed largely by capillarity (mean curvature flow) and potentially influenced by elasticity effects \cite{zhang2017,han2022}.

Modeling and simulations have been pivotal in understanding and tailoring single-crystalline SSD \cite{Kim2013,NaffoutiSCIADV2017,BollaniNC2019}. Studies on GB grooving and few-grain microstructures provided valuable insights \cite{Amran2014,Derkach2014,Derkach2017,Schiedung2017,Zigelman2023}, yet they remain limited in capturing the coupled GB migration and surface diffusion dynamics over many grains and long timescales. Phase-field (PF) models \cite{Chen2002,RATZ2006187,Steinbach_2009,LiCCP2009} have emerged as a leading framework for SSD studies. They realize an implicit description of geometries via smooth order parameter(s), which naturally handles topological changes, complex geometries, and three-dimensionality, all key aspects of SSD \cite{JiangACTAM2012,NaffoutiSCIADV2017,BollaniNC2019,BackofenIJNAM2019,GarckeJNS2023}. Convenient PF models coupling diffusion, phase transformation, and interface migration have been proposed \cite{Plapp2011,Choudhury2012,Schiedung2017,HoffroggePRE2021,MUKHERJEE2022111076}. 
Recent work has examined interesting aspects of polycrystalline thin film evolution, including the impact of thermal grooving on grain growth and its subsequent stagnation \cite{Verma2021,Verma2023}. First studies of polycrystalline dewetting have also been proposed \cite{Gao2022}, although with an approximated treatment of evolving morphologies and surface diffusion. An up and coming PF formulation, namely the grand-potential multi PF model \cite{Plapp2011,Choudhury2012,Zhang2025} (hereafter MPF), offers a general descriptions capturing the dynamics of both conserved and non-conserved order parameters, thus ideal for polycrystalline SSD scenarios which include (surface and bulk) diffusion as well as interface (GB) migration. Although this scenario has not been explored in detail, a recent implementation promisingly showed key phenomenology concerning thermal grooving at one GB \cite{HoffroggePRE2021}.

We present here the application of this MPF model to studying SSD in polycrystalline thin films. After outlining the model’s key features, we demonstrate its capabilities. We examine SSD onset through numerical simulations and benchmark its outcome against newly derived equations that predict critical grain aspect ratios and related features. Finally, we apply the MPF model to prototype polycrystalline films and to the relevant case of patches, highlighting similarities and differences with single-crystalline counterparts.

We employ the approach introduced in \cite{HoffroggePRE2021}, adapted here to systems dominated by surface/interface diffusion and interface migration, while neglecting bulk diffusion. We also consider grains of the same pure material and neglect any elastic effects arising from GB migration. Order parameters $\{\phi_1(\mathbf{r}),\dots,\phi_N(\mathbf{r})\}$$\equiv$$\,\boldsymbol{\varphi}$ akin to smooth characteristic functions track different domains/phases. Global concentration fields for $K$ elements $c_i(\mathbf{r})$ account for material transport. They can be related to the composition of each phase ($c_{i\alpha}$) via interpolation $c_i(\mathbf{r})=\sum_{\alpha=1}^N \phi_\alpha(\mathbf{r}) c_{i\alpha}
$ (spatial dependence will be omitted for brevity). To describe the chemical interactions within the MPF system, we define a set of $K - 1$ independent chemical potential  fields, $\{\mu_1,\dots,\mu_{K-1}\}\equiv\boldsymbol{\mu}$, corresponding to the concentration fields, which distinguish between the solid film, the substrate, and the surrounding vacuum. The grand potential functional reads
\begin{equation}
\begin{split}
\Psi[\boldsymbol{\varphi},\nabla\boldsymbol{\varphi},\boldsymbol{\mu}]=&\int_\Omega \bigg[f_{\rm I}(\boldsymbol{\varphi},\nabla\boldsymbol{\varphi}) + \psi (\boldsymbol{\varphi},\boldsymbol{\mu}) \bigg] \mathrm{d}\mathbf{r},\\
\end{split}
\end{equation}
where
\begin{equation}
f_{\rm I}=\sum_\alpha^N \sum_{\beta=\alpha+1}^{N} \gamma_{\alpha\beta} \left(\epsilon \nabla \phi_\alpha \cdot \nabla \phi_\beta + \frac{16\mathcal{G}(\boldsymbol{\varphi}) \phi_\alpha \phi_\beta}{\pi^2\epsilon}  \right).
\end{equation}
$f_{\rm I}$ encodes the interfacial energy with $\epsilon$ the interface thickness, $\mathcal{G}(\boldsymbol{\varphi})=1$ if $\sum_\alpha \phi_\alpha=1$ and $\phi_\alpha \geq 0,\ \forall \alpha$, while $\mathcal{G}(\boldsymbol{\varphi})=\infty$ otherwise. 
The function $\psi(\boldsymbol{\varphi},\boldsymbol{\mu})$ corresponds to the grand-potential density
\begin{equation}
        \psi(\boldsymbol{\varphi}, \boldsymbol{\mu}) = \sum_{i=1}^{K-1}\Big[\sum_{\alpha=1}^N (f_{i\alpha}-\mu_ic_{i\alpha})\phi_\alpha\Big].
\end{equation}
We consider here a parabolic free energy density $f_{i\alpha} = A_i^{\rm chem}(c_{i\alpha} - c_{i\alpha}^{\text{eq}})^2$ depending on the equilibrium concentrations $c_{i\alpha}^{\text{eq}}$ and  constants $A_i^{\rm chem}$.

Interface diffusion can be described by $\partial_t c_i =  - \nabla \cdot \mathbf{J}_{i}$ and $\mathbf{J}_{i}$ material flux for the $i$-th element constrained at the interfaces.
Upon assuming quasi-equilibrium, a uniform chemical potential for the $i$-th species across different phases is considered, namely $\mu_i = \frac{\partial f_{i1}}{\partial c_{i1}}=\ldots=\frac{\partial f_{iN}}{\partial c_{iN}}$. A Legendre transformation can be performed to substitute the chemical potential as a fundamental variable governing time evolution $c_{i\alpha}(t)=\frac{\mu_i(t)}{2A_i}+c_{i\alpha}^{\text{eq}}$. Phase boundary migration is accounted for through the dynamics of $\phi_\alpha$, as in classical PF models, with $\Psi$ being the energy functional. The resulting model equations read
\begin{equation}
\begin{split}
\epsilon \partial_t{\phi}_\alpha =&\frac{1}{\tilde N}\sum_{\beta=1}^{N}m_{\alpha \beta}\left(\frac{\delta \Psi}{\delta \phi_\beta}-\frac{\delta \Psi}{\delta \phi_\alpha}\right), \\
\frac{1}{2A_i} \partial_t \mu_i=&\nabla \bigg[\sum_{j=1}^{K-1}\sum_{\alpha=1}^N \sum_{\beta = \alpha + 1}^N \frac{32}{\epsilon\pi^2}M_{\alpha\beta j}\phi_\alpha\phi_\beta\nabla\mu_j \bigg]\\
&- \sum_{\alpha=1}^Nc_{i\alpha}^{\text{eq}}\partial_t\phi_\alpha,
\label{eq:mpf}
\end{split}
\end{equation}
with $m_{\alpha \beta}$ and $M_{\alpha\beta}$ the mobility and diffusivity at the $\alpha \beta$-interface, respectively. The timescale will be expressed in arbitrary units. Sharp interface limits can be found in \cite{HoffroggePRE2021,Hoffrogge2024}. Simulations are obtained by numerically solving Eqs.~\eqref{eq:mpf} using a finite difference integration scheme, implemented in the modular software package PACE3D \cite{HOTZER20181}.
Periodic boundary conditions are applied in the in-plane directions, while no-flux Neumann conditions are imposed at the top and bottom boundaries.

\begin{figure}
    \centering
    \includegraphics[width=\linewidth]{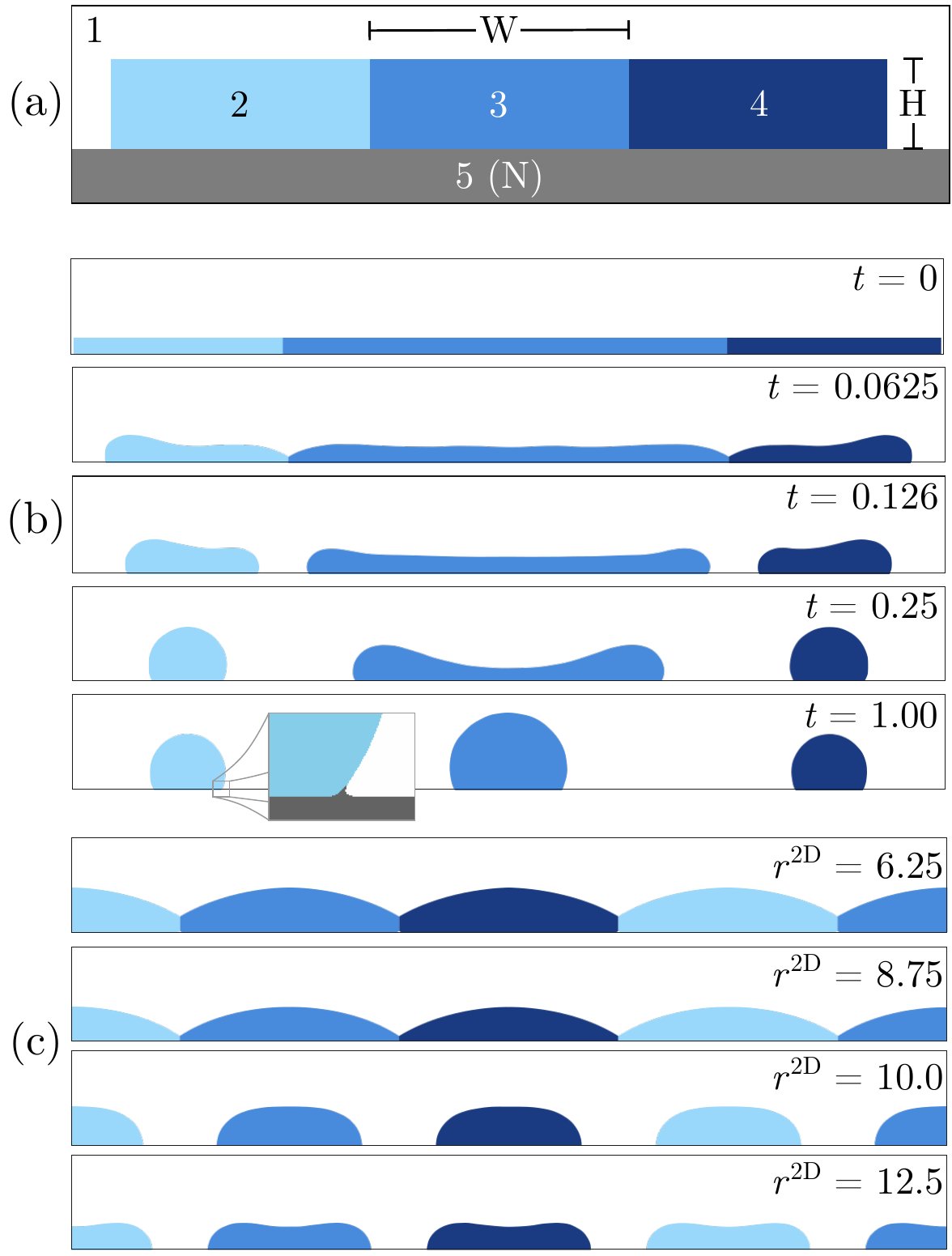}
    \caption{Modeling and 2D simulation of polycrystalline SSD. (a) Scheme of the considered geometries and labeling of the phases as considered in the MPF model. The grain aspect ratio is defined as $r^{\rm 2D}=W/H$. (b) SSD simulations ($\epsilon=0.2H$) of a thin film with three grains. The central one has an aspect ratio of $r^{\rm 2D}=25$. The two at the border of the film have both $r^{\rm 2D}=12.5$. The timescale is relative to the one needed to reach equilibrium. The inset shows the vapour-solid-substrate trijunction. (c) MPF simulations of SSD with varying $r^{\rm 2D}$ for a film ideally formed by grains of the same size. A critical aspect ratio of $r^{\rm 2D}_{\rm c} \approx 9$ is obtained in simulations.
    }
    \label{fig:figure1}
\end{figure}

\begin{figure}[t]
    \centering
    \includegraphics[width=\linewidth]{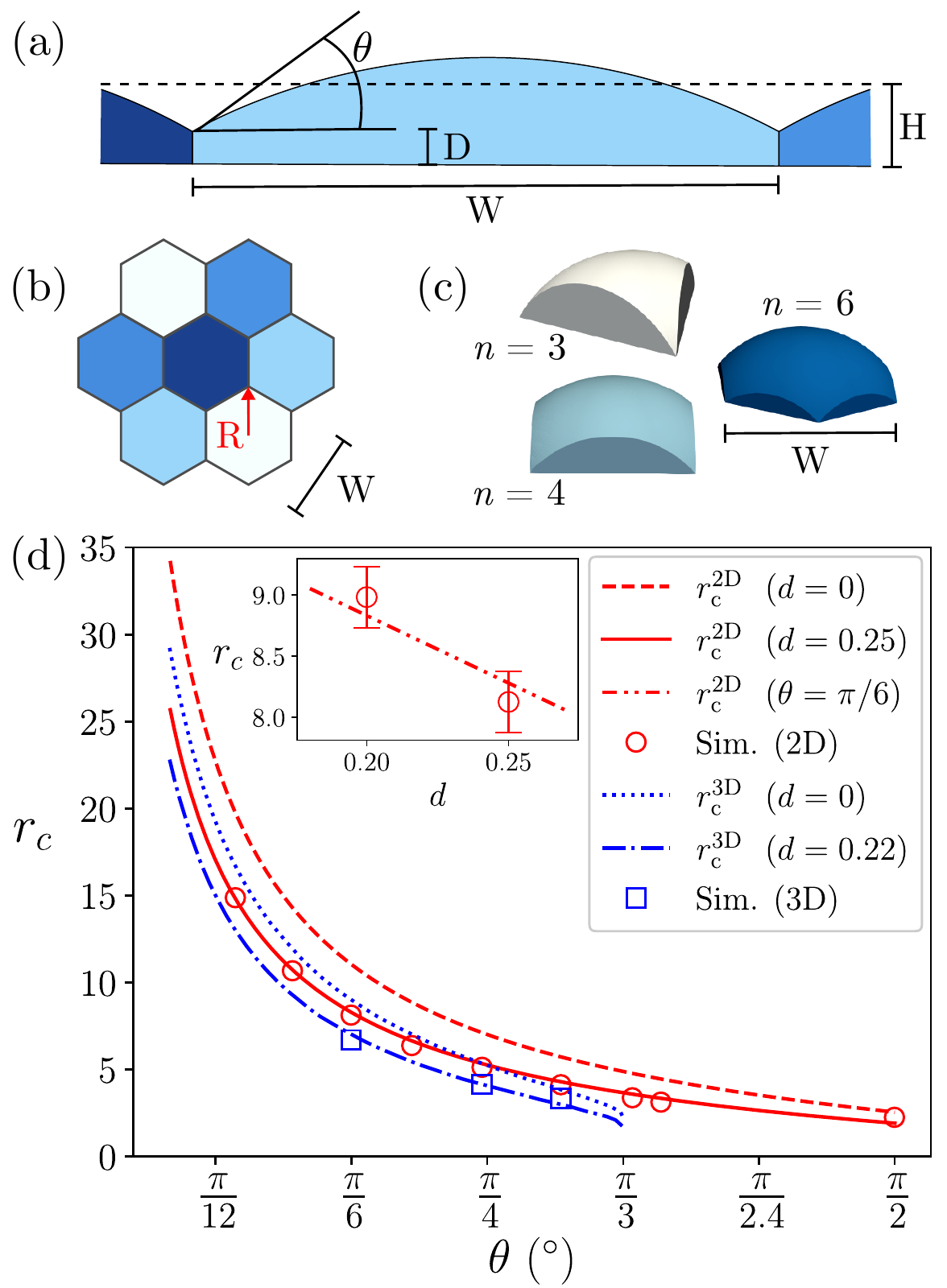}
    \caption{Geometrical parametrization and estimation of the critical aspect ratios. (a) Schematics of 2D equilibrium morphology. The dashed line illustrates the thickness ($H$) of the conformal initial film. (b) Schematics (top view) of the idealized setting of hexagonal grains in 3D with equal size. (c) Spherical caps truncated by $n$-gons. (d) $r_{\rm c}(\theta)$ in 2D and 3D as predicted by Eqs.~\eqref{eq:r2D} and ~\eqref{eq:r3D} for $d=0$ (sharp interface limit) and $d>0$ as in simulations. Symbols show $r_{\rm c}$ obtained by MPF simulations with $d=0.22$ (2D) and $d=0.25$ (3D). The inset illustrates the dependence of $r_{\rm c}$ on $d$ for $\theta=\pi/6$ (2D), as well as the average standard deviation of simulation results ($\pm 0.25$).}
    \label{fig:figure2}
\end{figure}

In our polycrystalline thin film model, the order parameters $\phi_\alpha$ track the substrate, the vapor phases, and the different grains within the film. Fig.~\ref{fig:figure1}a shows an illustration of a polycrystalline film composed of three grains, where 
we label the vacuum phase $\alpha=1$, the substrate $\alpha=N$, and the grains $1<\alpha<N$. We refer to the ratio of width to height $r^{\rm 2D}=W/H$ as the aspect ratio of the 2D grains. We use a two-concentration-field model for the substrate and the film with $c_{\text{substrate},N}^{\text{eq}}=1$ and $c_{\text{grain},\alpha}^{\text{eq}}=1$. We set $A_\text{substrate}^{\rm chem}=50$ and $A_\text{grain}^{\rm chem}=5$ \cite{Hoffrogge2024}. Interfacial (surface) diffusion is only allowed at the film-vapor interface via $M_{\alpha\beta}=\delta_{1\beta}$. We set $\gamma_{\alpha\beta}=0.5$ for all interfaces except otherwise specified. GB mobilities are set to $10^{-3}$ while all other mobilities are set to $1$. This choice allows for inspecting dewetting regimes with relatively slow interface motion. It does not prevent grain growth, as further illustrated below. Parameters match those used in Ref.~\cite{HOFFROGGE2023233031} for modeling (single-crystalline) nickel coarsening in solid oxide fuel cells. We indeed neglect interface-energy and mobility anisotropy, which are out of scope for the present discussion.

The simulation of SSD for a 2D three-grains film is shown in Fig.~\ref{fig:figure1}b. 
In the initial stages, rims form at the edges of the film as it retracts, while a groove develops at the GBs where the film eventually breaks. Afterward, each grain evolves according to the SSD dynamics of a single crystal. In the central grain at $t=0.25$, the typical material depletion in the center due to surface diffusion \cite{ThompsonARMR2012} can be observed, leading to a dent in the surface. For grains with large aspect ratios, as obtained after recrystallization, impingement of single-crystal patches, or abnormal grain growth, material depletion can cause breakpoints within grain centers, leading to more islands than initial grains.

\begin{figure*}
    \centering
    \includegraphics[width=\linewidth]{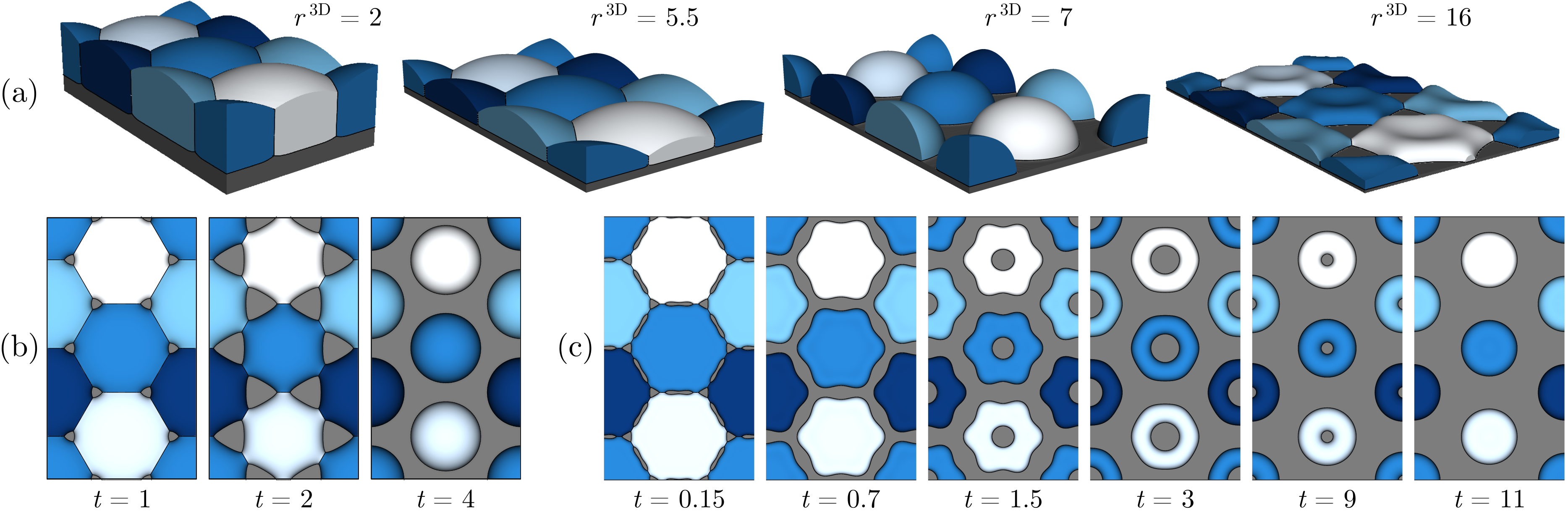}
    \caption{Simulations (3D) of morphological evolution and SSD in polycrystalline thin films with an idealized initial microstructure ($\epsilon=0.22H$). (a) Perspective view of the equilibrium configuration for thin films with varying aspect ratio, except for $r^{\rm 3D}=16$ for which an intermediate stage is reported. (b) Representative stages of SSD for grains with $r^{\rm 3D}=8$ (top view). The timescale is expressed relative to the breakup time. (c) Representative stages of SSD for grains with $r^{\rm 3D}=32$ (top view). The timescale is set as in panel (b).}
    \label{fig:figure3}
\end{figure*}

The grooves exhibit an (equilibrium) angle $\theta$ w.r.t the planar surface of the initial film that satisfies the identity $2\gamma_{\rm s}\sin \theta=\gamma_{\rm b}$ where $\gamma_{\rm s}$ and $\gamma_{\rm b}$ are the (isotropic) surface energy density of the crystal surface and the boundary, respectively. In this simulation $\gamma_{\rm s}=\gamma_{\rm b}=0.5$, i.e. $\theta=\pi/6$, well reproduced by the MPF simulations. For a GB between two infinitely extended grains ($r^{\rm 2D} \rightarrow \infty$), the depth of the grooves monotonously increases over time \cite{MullinsJAP1957}, thus inherently leading to breakup. 
For a finite aspect ratio, however, surface diffusion and thus grooving at the GBs stops when the equilibrium shape of the surface of the grain(s) is reached \cite{Srolovitz1986}. 
For isotropic surface energy, the equilibrium shape corresponds to a circular profile that satisfies the equilibrium angle $\theta$ at the groove. By simulating a periodic system with uniformly distributed grains and systematically increasing their aspect ratio, we identify a critical aspect ratio $r_{\rm c}^{\rm 2D}$ above which SSD occurs ($r^{\rm 2D}>r_{\rm c}^{\rm 2D}$). It results approximately $r_{\rm c}\approx 9$ for the chosen parameters, as illustrated in Fig.~\ref{fig:figure1}c. We remark that this occurs when the (diffuse) interface meets the substrate. 

Values of $r_{\rm c}^{\rm 2D}$ can be independently estimated by considering mass conservation \cite{Srolovitz1986}, i.e., equating the initial grain area to the area of a circular cap with contact angle $\theta$ and base $W$, i.e. $A_{\rm  G}=A_{\rm C}$, with $A_{\rm  G}=WH$ and $A_{\rm C}=W^2(2\sin\theta)^{-2}(\theta-\sin\theta \cos \theta)$. We also introduce an additional area $A_{D}=WD$ to account for breakup occurring when the groove reaches a distance $D$ from the substrate, namely in the presence of interfaces/surfaces with a finite effective thickness. The mass conservation condition thus reads $A_{\rm G}=A_{\rm C}+A_{\rm D}$. Expressing lengths as fractions of the initial film thickness ($W=rH$ and $D=dH$, implying $d<1$) and solving for the (critical) aspect ratio, we get
\begin{equation}
r_{\rm c}^{\rm 2D}=\frac{4\sin^2\theta}{\theta-\sin\theta\cos\theta}(1-d).
\label{eq:r2D}
\end{equation}
For $d=0$ and $\theta=\pi/6$ this results into $r_{\rm c}\approx 11$. By setting $d=\epsilon/H=0.2$, we get $r_{\rm c} \approx 8.8$, which is in excellent agreement with the simulation results. This holds upon varying $\theta$, which can be controlled by adjusting the solid–vapor interfacial energy ($\gamma_{\rm s}$, represented by $\gamma_{\alpha1}$ in the MPF model), and $d$, as illustrated in Fig.~\ref{fig:figure2}d.

For an idealized 3D case of uniformly sized, regular $n$-gonal grains perfectly tiling the substrate \cite{Srolovitz1986,MUKHERJEE2022111076} (Fig.~\ref{fig:figure2}b, $n=6$), the aspect ratio is defined as $r^{\rm 3D}=W/H$ where $W=2a$ and $a$ is the apothem of the $n$-gon. The volume of each grain is given by $V_{\rm G}=A_nH$ with the base area $A_n = n\tan(\pi/n)W^2/4$. The system is assumed to evolve toward a spherical cap truncated by the $n$-gon (see Fig.~\ref{fig:figure2}c). As derived in the Supplementary Material (SM), the corresponding integral expression for its volume is
\begin{equation}
V_{\rm C}
= \frac{nW^3}{4}\int_0^{\tilde{R}} \tilde{z}_n(\theta, \tilde{\rho}) \vartheta_n(\tilde{\rho}) \tilde{\rho} \,\mathrm{d} \tilde{\rho},\\ 
\end{equation}
with
\begin{equation}
\tilde{z}_n(\theta,\tilde{\rho}) = \sqrt{\sin^{-2}(\theta)-\tilde{\rho}^2}-\sqrt{\sin^{-2}(\theta)-\tilde{R}^2},
\end{equation}
$\tilde{\rho}=\rho/a$, $\tilde{R} =R/a= 1/\cos(\pi/n)$, $\vartheta_n(\tilde{\rho}) = \pi/n$ if $\tilde{\rho} < 1$ and $\pi/n - \arccos(\tilde{\rho}^{-1})$ otherwise.
By imposing mass conservation, $V_{\rm G}=V_{\rm C}+V_{\rm D}$ with $V_{\rm D}=A_nD$ and $D=dH$ in analogy to Eq.~\eqref{eq:r2D}, we then find the critical aspect ratio
\begin{equation}
r_{\rm c}^{\rm 3D}= \frac{\tan(\pi/n)}{\tilde{V}_{\rm C}(n, \theta)}(1-d).
\label{eq:r3D}
\end{equation}
It increases with $n$, indicating that grains with fewer edges undergo SSD earlier, suggesting that sharp corners promote film rupture. From the equations above, $\tilde{z}=0$ at the junction where $\tilde{\rho} = \tilde{R}$; since the (spherical) elevation decreases with radius, enhanced grooving at junctions, observed in previous works \cite{GENIN19923239,MUKHERJEE2022111076}, is naturally captured. The elevation of the junction (J) and of the GB at the midpoint between junctions can be expressed as (see SM)
\begin{equation}\label{eq:dj}
\begin{split}
d_{\rm J} &= H -  z_n(\theta, R), \\
    d_{\rm GB} &= \frac{W}{2}\left(\frac{1}{\tan{\theta_c}}-\sqrt{\frac{1}{\sin^2{\theta_c}}-\frac{1}{\cos^2({\pi}/{n})}}\right) + d_{\rm J}.   
\end{split}
\end{equation}

\begin{figure*}
    \centering
    \includegraphics[width=\linewidth]{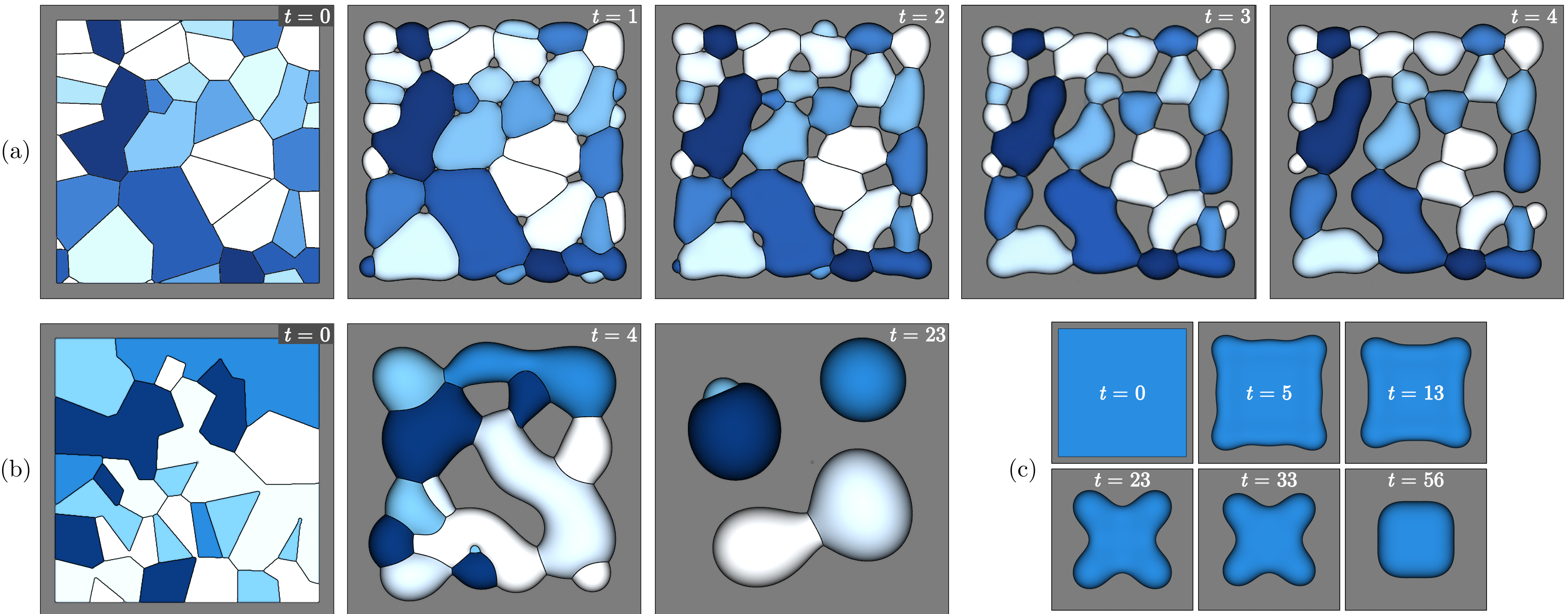}
    \caption{SSD of square polycrystalline patches with an arbitrary internal grain structure. The two panels (a) and (b) show representative stages of the SSD with two different initial conditions obtained through a Voronoi construction over randomly generated points. The resulting domains are then randomly assigned to different phase fields. These initial conditions show significantly different variability of the numbers of grains per side, which are $\sim 7.25 \pm 0.69$ in (a) and $\sim 4.75\pm 3.69$ (b). Timescale is given relative to the polycrystalline patch breakup time (similar in both cases). Panel (b) shows the evolution up to the late stages. (c) SSD of a single-crystalline patch with the same size and aspect ratio as in (a) and (b).
    }
    \label{fig:figure4}
\end{figure*}

MPF three-dimensional simulations of SSD of the idealized initial configurations in Figs.~\ref{fig:figure2}b are presented in Fig.~\ref{fig:figure3}a, varying $r^{\rm 3D}$. For relatively small aspect ratios, grain morphologies at equilibrium are well described by hexagonally truncated spherical caps ($n=6$, Fig.~\ref{fig:figure2}c). Both $d_{\rm J}$ and $d_{\rm GB}$ decrease with increasing $r^{\rm 3D}$ and agree with Eq.~\eqref{eq:dj}. For instance, at $r^{\rm 3D}=5.5$, simulations give
$d_{\rm J}/H \approx 0.37$ and $d_{\rm GB}/H \approx 0.65$, in almost perfect agreement with predictions of Eq.~\eqref{eq:dj}, namely $d_{\rm J}/H \approx 0.37$ and $d_{\rm GB}/H \approx 0.64$. The proposed Eq.~\eqref{eq:r3D} further estimates $r_{\rm c}^{\rm 3D}\approx8.7$ for $d=0$ and $r_{\rm c}^{\rm 3D}\approx6.8$ for $d=\epsilon/H=0.22$, well matching the simulated critical aspect ratio $r_{\rm c}^{\rm 3D}\approx6.7$  (Fig.~\ref{fig:figure3}a). This agreement holds for other $\theta$ values too, as illustrated in Fig.~\ref{fig:figure2}d. 
The MPF framework is thus benchmarked against theoretical predictions from Eqs.~\eqref{eq:r3D} and \eqref{eq:dj}, which also independently enable the estimation of the SSD onset in polycrystalline thin films from the grain aspect ratio. We note, however, that localized inhomogeneities and defects may trigger SSD with pronounced heterogeneity.

Figure~\ref{fig:figure3}b illustrates the morphological evolution during dewetting described by MPF simulations, considering an idealized initial configuration as in Fig.~\ref{fig:figure2}b and a relatively large $r^{\rm 3D}$ to ensure film rupture. SSD initiates at junctions, confirming seminal equilibrium considerations \cite{Srolovitz1986}, other recent simulations \cite{MUKHERJEE2022111076}, and quantitative predictions discussed above. Additionally, a detailed account of the dynamics at GB junctions is obtained, including the formation and growth of non-trivial opening holes. At intermediate stages, grains detach from neighbors and evolve into single-crystalline patches. A similar qualitative behavior, enriched by additional features, emerges at larger aspect ratios; see Fig.~\ref{fig:figure3}c. Importantly, breakup is found to initiates at triple junctions and rapidly propagates along grain boundaries, forming multiple contact points. When the grains detach completely from each other, the dynamics reduces to single-crystalline SSD. A rim forms at the edges (see also Fig.~\ref{fig:figure3}a for $r^{\rm 3D}=16$). For the considered aspect ratio, a breakup at the center of the patch is obtained, resembling a classical feature of single-crystalline SSD~ \cite{YeADVMAT2011, JiangACTAM2012, NaffoutiSCIADV2017}. A semi-toroidal shape forms and, driven by curvature gradients, evolves until the central hole eventually closes.

We finally present a novel scenario: SSD of a polycrystalline patch; see Figs.~\ref {fig:figure4}a and \ref {fig:figure4}b illustrating the SSD for two patches with varying grain morphologies and aspect ratios. Grooves form at GBs, while holes open at junctions. Such holes initiate dewetting fronts within the patch, while a continuous rim effectively develops along the edges, breaking only at later stages. Aside from the rapid disappearance of small grains, the slower dynamics of rims and peripheral GBs, compared to hole opening, are evident along edges with both relatively small and large grains. We thus find that SSD in polycrystalline patches leads to early-stage morphologies characterized by material accumulation between the center and the initial patch boundaries, closely resembling the behavior observed in single crystals \cite{JiangACTAM2012, NaffoutiSCIADV2017}, reported in Fig.~\ref{fig:figure4}c for completeness. In typical polycrystals, the presence of GBs leads to random, uncontrolled breakup within the patch. However, designing GB networks can provide additional degrees of freedom to control self-assembly. Notably, hole opening occurs much faster than rim retraction driven by surface diffusion (c.f., Fig.~\ref{fig:figure4}).

In conclusion, we demonstrated a grandpotential MPF model tackling the geometric complexity of SSD dynamics in polycrystalline thin films. The simulation results are consistent with novel independent predictions based on energetic arguments and  characterize the central, non-trivial morphological evolutions. Expressions of the grain critical aspect ratio for dewetting have been derived and represent a theoretical benchmark for scenarios well beyond the simulations reported here. We further elucidated key aspects of the dewetting onset and analyzed the SSD of polycrystalline patches, highlighting the differences and similarities with single-crystalline SSD. Tailoring GB distributions can expand the design space for SSD-driven self-assembly. Our approach provides a versatile tool for exploring the associated parameter space. For the sake of generality, we have illustrated idealized, minimal settings, particularly with respect to material parameters. Nevertheless, such simplified conditions have already proven powerful for understanding and predicting single-crystalline solid-state dewetting \cite{JiangACTAM2012,NaffoutiSCIADV2017}, and we reported here for the first time the characteristic traits of their polycrystalline counterparts. The considered MPF model still allows for extended parameterizations and couplings compatible PF models,  e.g., incorporating anisotropic surface or GB energy densities \cite{torabiPRSA2009,SalvalaglioCGD2015,Ribot_2019}, elasticity, and plasticity effects \cite{RATZ2006187,Higgins2021,Qiu2025}.
Careful assessment against theoretical predictions and experimental observations will be essential to quantitatively assess the influence of these additional contributions. Incorporating specifically shear-coupled GB migration \cite{Qiu2025} still requires extending the underlying continuum model to fully 3D geometries and accounting for free surfaces of arbitrary morphology. While these remain open challenges, the approach proposed here provides a solid foundation for future developments that fully incorporate capillarity and capture concurrent kinetic processes.

\subsection*{Data Availability}

The data supporting the findings of this article are openly available \cite{becker2025solid}.

\subsection*{CRediT authorship contribution statement}
\textbf{Paul Hoffrogge}: Data curation (equal), Investigation (equal), Methodology (equal), Software (equal), 
Visualization (equal), Writing – review \& editing (equal). 
\textbf{Nils Becker}: Investigation (equal), Data curation (equal), Methodology (equal), Software (equal), Visualization (equal), Writing – review \& editing (equal). 
\textbf{Daniel Schneider}: Software (support), Data curation (equal), Funding acquisition (equal), Methodology (support), Supervision (support). \textbf{Britta Nestler}: Software (support), Funding acquisition (equal), Supervision (lead)
\textbf{Axel Voigt}: Conceptualization (support), Funding acquisition (equal).
\textbf{Marco Salvalaglio}: Conceptualization (lead), Investigation (support), Funding acquisition (equal),
Methodology (support),
Supervision (support),
Visualization (equal), Writing – original draft, Writing – review \& editing (equal)

\subsection*{Declaration of Competing Interest}

The authors declare that they have no known competing financial interests or personal relationships that could have appeared to influence the work reported in this paper.

\subsection*{Acknowledgements}

This work was funded by the German Research Foundation (DFG) under Project ID 390874152 (POLiS Cluster of Excellence). M.S. and A.V. acknowledge support from the Deutsche Forschungsgemeinschaft (DFG, German Research Foundation) within FOR3013, project number 417223351. D.S. and B.N. are grateful for support from the Helmholtz association within programme MSE (P3T1) number 43.31.01. The authors gratefully acknowledge the computing time made available to them on the high-performance computer at the NHR Center of TU Dresden. This center is jointly supported by the Federal Ministry of Education and Research and the state governments participating in the NHR.

\clearpage
\newpage

\onecolumngrid

\begin{center}
  \textbf{\large \hspace{5pt} SUPPLEMENTARY MATERIAL \\ \vspace{0.2cm} Solid-State Dewetting of Polycrystalline Thin Films: a Phase Field Approach \\}
\vspace{0.4cm}   Paul Hoffrogge,$^{1,2,*}$, Nils Becker,$^{3,*}$, Daniel Schneider,$^{1,2,3}$\\ Britta Nestler,$^{1,2,3}$ Axel Voigt,$^{4,5}$ and Marco Salvalaglio$^{4,5,\dagger}$ \\[.1cm]
  {\itshape \small
  ${}^1$Institute of Nanotechnology—Microstructure Simulations (INT-MS),\\ Karlsruhe Institute of Technology, Hermann-von-Helmholtz-Platz 1, 76344
  \\ 
  ${}^2$Institute for Digital Materials Science (IDM), \\ Karlsruhe University of Applied Sciences, Moltkestrasse 30, 76133 Karlsruhe, Germany
 \\
  ${}^3$Institute for Applied Materials (IAM-MMS),\\ Karlsruhe Institute of Technology, Strasse am Forum 7, 76131 Karlsruhe, Germany
 \\
  ${}^4$Institute  of Scientific Computing,  TU  Dresden,  01062  Dresden,  Germany
   \\
  ${}^5$Dresden Center for Computational Materials Science (DCMS), TU Dresden, 01062 Dresden, Germany
 } \\
  \vspace{0.1cm}{\small
   ${}^*$ These authors contributed equally to this work\\
   ${}^\dagger$ marco.salvalaglio@tu-dresden.de
}
\vspace{0.5cm}
\end{center}

\twocolumngrid

\setcounter{equation}{0}
\setcounter{figure}{0}
\setcounter{table}{0}
\setcounter{section}{0}
\setcounter{page}{1}

\renewcommand{\thesection}{S-\Roman{section}}
\renewcommand{\theequation}{S-\arabic{equation}}
\renewcommand{\thefigure}{S-\arabic{figure}}
\renewcommand{\bibnumfmt}[1]{[S#1]}
\renewcommand{\citenumfont}[1]{S#1}

\section*{Theory of N-gonal Polycrystalline Thin-Films}

\subsection*{Angles and base area}

\begin{figure}[!ht]
    \centering
    \hfill
    \begin{minipage}[t]{0.22\textwidth}
        \centering
        \includegraphics[scale=1]{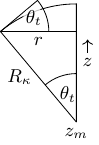}\\
        (a) Side view
    \end{minipage}
    \hfill
    \begin{minipage}[t]{0.22\textwidth}
        \centering
        \includegraphics[scale=1]{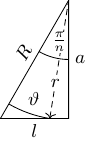}\\
        (b) Top view
    \end{minipage}
    \hfill
 \caption{Spherical cross-section (a) and $n$-gon unit cell (b) of a grain in the tile pattern.}
    \label{fig:grain_views}
\end{figure}

A grain with a spherical surface, having radius of curvature $R_\kappa$ as depicted in Fig.~\ref{fig:grain_views}a is assumed. The tangent angle $\theta_t$ is related to the radius of curvature as follows:
\begin{equation}\label{eq:eq1SM}
 r = R_\kappa \sin(\theta_t).
\end{equation}
Let us consider the grain to be truncated by a regular $n$-gon such that the grain is in contact with its $n$ neighbor grains at a distance, named the apothem $a$ of the $n$-gon. Here, we assume a constant contact angle $\theta$ is realized with all its neighbors, giving
\begin{align}
 a = R_\kappa \sin(\theta),
\end{align}
which then, using Eq.~\eqref{eq:eq1SM}, implies
\begin{align}
  \theta_t(r) = \arcsin\left(r\sin(\theta)/a \right).
\end{align}
The outerscribed circle radius of the $n$-gon is denoted as $R$ (see Fig.~\ref{fig:grain_views}b). The area of the $n$-gons unit cell, is given by $al/2$, thus the total base area of the grain is $A_n = nal$ and since $l = a \tan(\pi/n)$, we have
\begin{align}
  A_n = na^2 \tan(\pi/n).
\end{align}
This triangle also reveals the relation
\begin{align}
 a &= R \cos(\pi/n).
\end{align}
The vertices of the $n$-gon correspond to multi-junctions. The tangent angle of the surface at the junction is given by
\begin{align}
 \theta_{\mathrm{junction}} = \theta_{t}(R) = \arcsin\left(\frac{\sin(\theta)}{\cos(\pi/n)}\right).
\end{align}
The condition $\theta_{\mathrm{junction}} < \pi/2$ must be satisfied for the junction to exist; otherwise, immediate dewetting occurs. Therefore,
\begin{align}
 \sin(\theta) < \cos(\pi/n) = \sin(\pi/2\pm\pi/n),
\end{align}
is obtained, which gives the maximal contact angle
\begin{align}
\theta_{\mathrm{max}} = \pi/2-\pi/n\,.
\end{align}

\subsection*{Volume of N-gonal grain}
Let us assume $\theta < \theta_{\mathrm{max}}$, then the elevation of the surface $z_n$ of the $n$-gon is, according to Pythagorean theorem (see Fig.~\ref{fig:grain_views}a),
\begin{align}
(z_n(\theta, r)-z_m(\theta))^2 + r^2 = R_\kappa^2(\theta),
\end{align}
and hence
\begin{align}
z_n(\theta, r) = z_m + \sqrt{R_\kappa^2(\theta) - r^2}.
\end{align}
The elevation at which the junction pit forms is given as $z_n(\theta, R)$. Within the grain, the elevation is lowest at the junction, hence we enforce without loss of generality $z_n(\theta, R) = 0$ to obtain elevations relative to the junction pit.
This yields
\begin{align}
z_m = -a\sqrt{\sin^{-2}(\theta) - \cos^{-2}(\pi/n)}.
\end{align}

The volume $V_{\rm C}$ of the $n$-gon grain featuring a spherical cap can be calculated by integration of the elevation in cylindrical coordinates:
\begin{align}
V_{\rm C} = 2n V_{\text{unit cell}} = 2n\int_{0}^{R} z_n(\theta, \rho) \vartheta_n(\rho) \rho \, \mathrm{d} \rho,
\end{align}
where $\vartheta_n(\rho)$ is the angle between two edges of the $n$-gon unit cell (see Fig.~\ref{fig:grain_views}b),
\begin{align}
\vartheta_n(\rho) = \begin{cases} \pi/n & \rho < a \\ \pi/n - \arccos(a/\rho) & \rho \geq a\end{cases}\,.
\end{align}
Introducing the dimensionless quantities $\tilde{z}_n = z_n / a$, $\tilde{R} = {R} / a$, $\tilde{\rho} = \rho/a$ and $\tilde{V}_{\rm C} = V_{\rm C} /(2na^3)$, we can rewrite
\begin{align}
V_{\rm C} = 2na^3\tilde{V}_{\rm C}(\theta) = 2na^3\int_{0}^{\tilde{R}} \tilde{z}_n(\theta, \tilde{\rho}) \vartheta_n(\tilde{\rho}) \tilde{\rho} \, \mathrm{d} \tilde{\rho},
\end{align}
where 
\begin{align}
 \tilde{R} &= 1/\cos(\pi/n),\\
 \tilde{z}_n(\theta, \tilde{\rho}) &= \sqrt{\sin^{-2}(\theta) - \tilde{\rho}^2} - \sqrt{\sin^{-2}(\theta) - \tilde{R}^2},\\
\vartheta_n(\tilde{\rho}) &= \begin{cases} \pi/n & \tilde{\rho} < 1 \\ \pi/n - \arccos(\tilde{\rho}^{-1}) & \tilde{\rho} \geq 1\end{cases}\,.
\end{align}

\subsection*{Critical aspect ratio}

Let us consider an initially flat $n$-gonal grain with height $H$. Its volume is then given by
\begin{align}
V_{\rm G} = A_nH.
\end{align}
Let us assume that the grain coarsens, eventually reaching the stable quasi-equilibrium truncated spherical shape (see also the main text, Fig. 2c). It is assumed that the film breaks up at some height $D < H$, giving the breakup volume
\begin{align}
V_{\text{breakup}} &= V_{\rm C} + A_nD.
\end{align}
Therefore, the breakup is ensured at a finite time when the volume of the grains satisfies the following condition
\begin{equation}
\begin{split}
V_{\rm G} &< V_{\text{breakup}},
\\
A_nH &< 2na^3\tilde{V}_{\rm C}(\theta) + A_nD,
\\
\frac{A_n}{2na^3} &< \frac{\tilde{V}_{\rm C}(\theta)}{H-D}.
\end{split}
\end{equation}
Since the function $1/x$ is strictly montonously decreasing for $x > 0$ it follows
\begin{align}
\frac{2na^3}{A_n} &> \frac{H-D}{\tilde{V}_{\rm C}(\theta)}.
\end{align}
With $D$ expressed in terms of the film thickness, $D=dH$, with $d<1$ an adimensional factor, we have
\begin{align}
\frac{2na^3}{A_nH} > \frac{1-d}{\tilde{V}_{\rm C}(\theta)},
\end{align}
and with $A_n = na^2 \tan(\pi/n)$ we obtain
\begin{align}
\frac{2a}{H\tan(\pi/n)} > \frac{1-d}{\tilde{V}_{\rm C}(\theta)}.
\end{align}
The critical aspect ratio $r^{\rm 3D}_{\rm c}$, defined as twice the apothem divided by the initial height $H$ of the film, $r^{\rm 3D}=2a/H$ (see also the main text), thus results
\begin{align}
r^{\rm 3D}_{\rm c}= \frac{2a}{H} = \frac{\tan(\pi/n)}{\tilde{V}_{\rm C}(\theta)}(1-d).
\end{align}

\subsection*{Relative Depth of GB and Junction}
The depth of the GB and junction w.r.t. the initial film height is given by $d_{\rm GB} = H - z_n(\theta, a)$ and $d_{\rm J} = H -  z_n(\theta, R)$ when we compare equal volumes (i.e. $d=0$).
The relative depth of the junction and GB can then be computed as follows
\begin{equation}\label{eq:dratio}
\begin{split}
\frac{d_{\rm J}}{d_{\rm GB}} &= \frac{H -  z_n(\theta, R)}{H - z_n(\theta, a)} = \frac{H/a}{H/a - \tilde{z}_n(\theta, 1)}
\\
 &= \frac{1}{1 - \frac{\tan(\pi/n)}{2\tilde{V}_{\rm C}(\theta)}\tilde{z}_n(\theta, 1)}.
 \end{split}
\end{equation}

When the film is initially flat, we expect at early times that the groove progresses almost ideally following the early prediction reported in Ref.~\cite{MullinsJAP1957}, with the grain interior acting as a half-infinite space. Thus, the depth of the groove is approximately (for small $\theta$) $d_{\rm GB}(t) \approx 0.78\tan(\theta)(Bt)^{1/4}$ where $B$ is the rate constant for surface diffusion.
When $r^{\rm 3D} \gg r_{\rm c}^{\rm 3D}$, rupture occurs at early times, where this can be considered a good approximation. Assuming that rupture initiates at the junction, as evidenced also in the main text, and that the relative depth ratio derived approximates also dynamic junctions well, Eq.~\eqref{eq:dratio} can be used to estimate the critical rupture time, thus reading $t_{\rm C} = H^4/B(0.78\tan(\theta)d_{\rm J}/d_{\rm GB})^{-4}$. When $r^{\rm 3D} \gtrsim r_{\rm c}^{\rm 3D}$ rupture may occur later than $t_{\rm C}$ as the finite grain size slows down the grooving dynamics.

\end{document}